\documentclass[11pt]{article}
\usepackage{amssymb}
\usepackage{amsmath}
\usepackage[all]{xy}
\input{epsf}
\usepackage{epsfig}
\usepackage{hyperref}
\numberwithin{equation}{section}

\usepackage{color}

\newcommand{\bea}{\begin{eqnarray}}
\newcommand{\eea}{\end{eqnarray}}
\def\be{\begin{equation}}
\def\ee{\end{equation}}
\def\ba{\begin{align}}
\def\ea{\end{align}}
\def\beq{\begin{eqnarray}}
\def\eeq{\end{eqnarray}}

\begin{document}

\title{The fate of Newton's law in brane-world scenarios}

\author{Raphael Benichou $^\flat$ and John Estes $^\natural$}

\maketitle

\begin{center}
$^\flat$ Theoretische Natuurkunde, Vrije Universiteit Brussel and \\
The International Solvay Institutes,\\
Pleinlaan 2, B-1050 Brussels, Belgium \\
\textsl{raphael.benichou@vub.ac.be}
\end{center}

\vskip 3mm

\begin{center}
$^\natural$ Instituut voor Theoretische Fysica, Katholieke Universiteit Leuven, \\
Celestijnenlaan 200D B-3001 Leuven, Belgium \\
\textsl{johnalondestes@gmail.com}
\end{center}

\begin{abstract}
We consider brane-world scenarios embedded into string theory.
We find that the D-brane backreaction induces a large increase in the open string's proper length.
Consequently the stringy nature of elementary particles can be detected at distances much larger than the fundamental string scale.
As an example, we compute the gravitational potential between two open strings ending on backreacting D3-branes in four-dimensional compactifications of type II string theory.
We find that the Newtonian potential receives a correction that goes like $1/r$ but that is not proportional to the inertial masses of the open strings, implying a violation of the equivalence principle in the effective gravitational theory.
This stringy correction is screened by thermal effects when the distance between the strings is greater than the inverse temperature.
This suggests new experimental tests for many phenomenological models in type II string theory.
\end{abstract}

\maketitle

\newpage

D-branes are non-perturbative extended objects which play a key role in many recent developments of string theory.
The elementary degrees of freedom of D-branes are open strings with endpoints attached to the D-branes.
String theory has been extensively studied in the probe-brane approximation, while
much less is understood when the backreaction of the D-branes is taken into account.
In this letter we compute the gravitational potential energy between open strings ending on backreacting D3-branes.
We find that the D-brane backreaction yields a large correction to the gravitational potential.
This happens because the gravitons exchanged between the open strings are sensitive to the geometry in the neighborhood of the D-branes, which is strongly deformed by the D-brane backreaction.

Our results are relevant for a large family of phenomenological models,
which are promising candidates to embed our universe into string theory.
In these scenarios, the standard model is localized on D-branes
arranged in a four-dimensional compactification of superstring theory.
Our computations show that an observer living on such a braneworld could detect a macroscopic violation of the equivalence principle at very low temperatures.


\section{The setup}

\begin{figure}[t]
\centering
\includegraphics[width=1\linewidth]{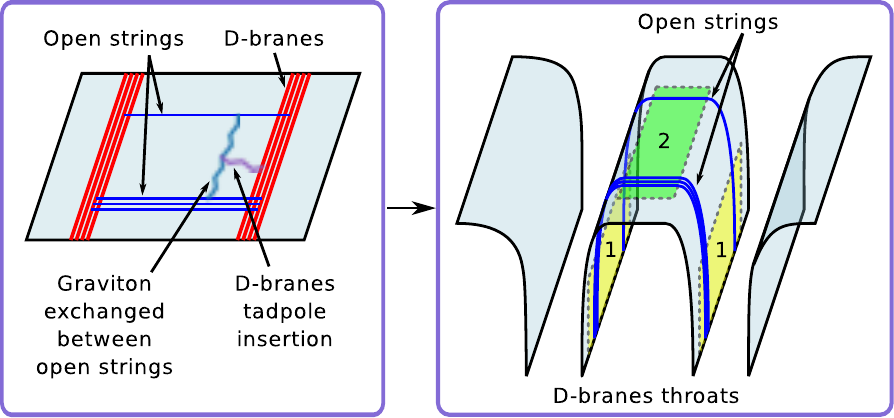}
\caption{\emph{Left:} Open strings stretched between two stacks of D-branes. The D-brane backreaction affects the gravitational interactions between the open strings.
\emph{Right:} To take into account the  full D-brane backreaction we work in the D-brane background.
Most of the inertial mass of the strings is located in region 2. However we show that a non-negligible contribution to the gravitational potential comes from regions 1.}
\label{figBackreact}
\end{figure}

We first consider the case of open strings ending on probe D-branes in a flat 10-dimensional spacetime.  When the separation of the strings is much larger than their length, the gravitational potential goes like $m_1 m_2 r^{-7}$, where $r$ is the distance between the two strings. The inertial masses of the strings $m_{1,2}$ are given by the product of the string tension times the string length.  The probe approximation is exact at zero string coupling. For non-zero string coupling one must include the backreaction of the D-branes, which will modify the gravitational potential.  In a perturbative approach, the D-branes backreaction is encoded in the form of tadpole insertions in the graviton exchange diagrams (Figure \ref{figBackreact}, \emph{left}).  In this letter we take into account the full D-brane backreaction, by evaluating the gravitational potential between open strings in the D-brane background (Figure \ref{figBackreact}, \emph{right}).

We focus on the case of D3-branes which are obvious candidates for model building \footnote{Our arguments are rather generic and our results should qualitatively generalize to the cases of other Dp-branes.}.
The geometry resulting from the backreaction of a stack of D3-branes in ten-dimensional flat space is:
\be\label{metricD3} ds^2 = H(u)^{-\frac{1}{2}} dx_\mu dx^\mu + H(u)^{\frac{1}{2}}\left(du^2 + u^2 d\Omega_{(5)}^2\right) \ee
where $H(u) = 1+L^4/u^4$ and $L$ is given by $L^2 = \sqrt{4\pi g_s N}\alpha'$ with $N$ the number of D3-branes, $g_s$  the closed string coupling and $\alpha'$ the inverse string tension.

We consider open strings in the geometry \eqref{metricD3}. These open strings extend up to the horizon at $u=0$ where the D3-branes are sitting. For simplicity we consider only straight strings that extend along the $u$ direction, and are localized in all other directions. The inertial mass of such open strings is given by the integral of the string tension along the string. For a string stretching from $u=u_0$ \footnote{We can think of this string as being attached to a probe D3 at $u=u_0$.} up to $u=0$, the inertial mass is simply $u_0 \alpha'$. Notice that the proper length of this string is infinite, while the inertial mass is finite thanks to the infinite redshift at the horizon.  From the point of view of particle physics, such string states could describe matter and massive gauge bosons of the standard model.

We are interested in computing the gravitational attraction between two such open strings in the D3-brane geometry \eqref{metricD3}.
The sections of the open strings which are located outside the D3-branes throat carry most of the inertial mass and they provide the usual Newtonian contribution to the gravitational potential $m_1 m_2 r^{-7}$.
We will show that the section of the strings located in the depths of the D3-branes' throat provides another non-negligible contribution.
To perform our computation we can focus on the near-horizon D3-brane geometry \footnote{We keep gravity coupled to the open string theory, contrary to what is usually done in the context of AdS/CFT}.
This  means that our results hold independently of the details of the asymptotic geometry.
Introducing the coordinate $z=L^2/u$, the geometry \eqref{metricD3} simplifies in the near-horizon region to $AdS_5 \times S^5$:
\be\label{metricAdS} ds^2= L^2 \frac{dz^2 -dt^2 + dr^2 + r^2 d\Omega^2_{(2)}}{z^2} + L^2 d\Omega^2_{(5)} \ee
The D3-brane worldvolume now coincides with the Poincar\'e horizon at $z=\infty$.


\section{Gravitational attraction between open strings ending on D3-branes}

We now compute the gravitational potential between open strings ending on the extremal D3-branes as described above.  This problem was first studied in \cite{Benichou:2010sj} using linearized Einstein gravity.
Here we perform the computation in the full non-linear ten-dimensional IIB supergravity theory.

\begin{figure}[t]
\centering
\includegraphics[width=1\linewidth]{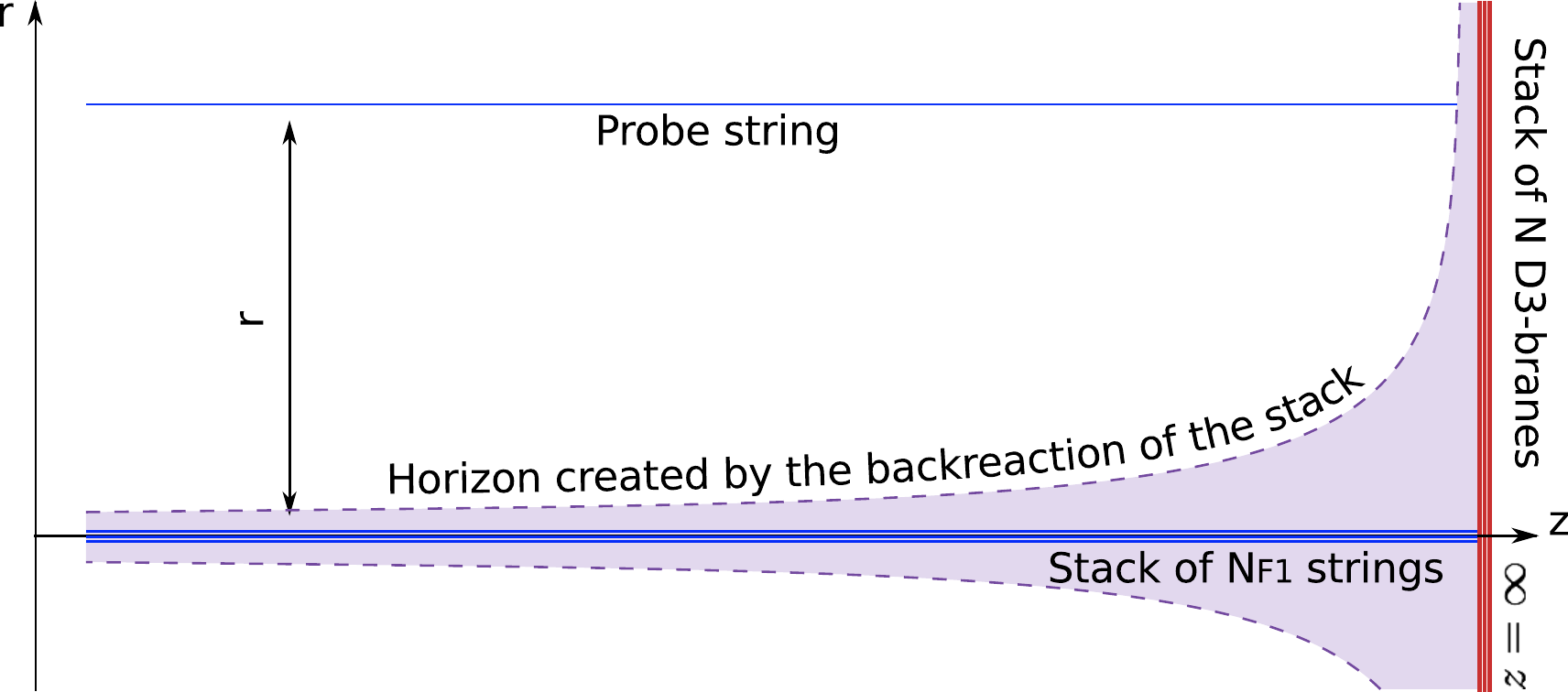}
\caption{ Open strings in the near-horizon of the D3-branes. The backreaction of the stack of strings effectively pulls out the horizon of the D3-branes.}
\label{fig1}
\end{figure}

We consider a stack of open strings located at $r=0$ in the geometry \eqref{metricAdS}. We compute the gravitational force that this stack applies on another string located at finite $r$ (see Figure \ref{fig1}).  The coordinate distance $r$ between the strings is the distance that an observer living on the D-branes would measure. For simplicity all strings are located at the same point of the five-sphere and we will compute only the leading term in the gravitational potential in the large $r$ limit.

To evaluate the gravitational potential energy between two objects, a general method is to evaluate the action for one object in the gravitational field created by the other object.
For instance to compute the potential energy between two point-like masses in flat space, one would evaluate the action for the first particle $S \propto\int ds$ in the Schwarzschild spacetime resulting from the gravitational backreaction of the second particle.  In our case, we evaluate the Nambu-Goto action for the single string placed in the supergravity solution corresponding to the backreaction of the stack of open strings.

The explicit supergravity solution describing a half-BPS stack of open strings in the near-horizon region of D3-branes was found in \cite{D'Hoker:2007fq}.
The geometry can be conveniently written in the form:
\be\label{metricWilsonLine} ds^2 = f_1^2 ds_{AdS_2}^2 + f_2^2 ds_{S^2}^2 +  f_4^2 ds_{S^4}^2 + 4 \rho^2 ds_\Sigma^2 \ee
where $\Sigma$ is a Riemann surface with disc topology. The functions $f_1$, $f_2$, $f_4$ and $\rho$ depend only on the coordinates parameterizing the Riemann surface $\Sigma$.
The $AdS_5\times S^5$ geometry \eqref{metricAdS} can be written in these coordinates as: $ds^2 = L^2(\cosh^2 (x) ds_{AdS_2}^2 + \sinh^2(x) ds_{S^2}^2 +  \cos^2(\theta) ds_{S^4}^2 + dx^2 + d\theta^2 )$. The Riemann surface $\Sigma$ is then the semi-infinite rectangle parametrized by $x$ and $\theta$\footnote{Writing the $AdS_2$ metric as $ds_{AdS_2}^2 = (dw^2-dt^2)/w^2$, we recover the parametrization \eqref{metricAdS} with the change of variables $z=w/\cosh(x)$, $r=w \tanh(x)$.}.  In the probe approximation, the stack of fundamental strings lies along the $AdS_2$ slice and sits at $x=0$ and $\theta = 0$.  Including the backreaction, deforms the warp factors and one finds that the topology of the space-time changes so that there are additional cycles supporting fundamental string charge.  We refer to \cite{D'Hoker:2007fq} and \cite{BE} for more details.

We evaluate the Nambu-Goto action for the single string in the geometry \eqref{metricWilsonLine}.
The worldsheet is extended in the time direction $t$. It remains to specify a path for the string in a constant-$t$ spacelike slice.
In the asymptotic anti-de Sitter region the string is located at constant $r$, and it extends up to the horizon.
The precise choice of path near the horizon does not influence qualitatively our results.
We simply assume that the path reduces to the straight line $r=cst$ when the backreaction of the stack vanishes.

The Nambu-Goto action can be written as $ S_{NG} = -(m+E(r)) (\int dt )$,
where $m$ is the inertial mass of the string and  $E(r)$ is the gravitational potential. For large $r$, we find:
\be\label{energy(r)} E(r) = -  \frac{\alpha'^4 g_s^2}{L^6} \frac{N_{F1}\sqrt{g_s N}}{\alpha'} \frac{1}{{r}} + {\cal O}\left(\frac{1}{r^2}\right) \ee
where ${\alpha'}^4 g_s^2$ is the ten-dimensional Newton constant and $N_{F1}$ is the number of fundamental strings in the stack. We omitted an overall numerical factor that depends on the choice of path and can be evaluated numerically.
Details of this computation are given in \cite{BE}.
The $1/r$ behavior of the potential \eqref{energy(r)} can be understood from a linearized gravity analysis \cite{Benichou:2010sj}.

The gravitational potential energy \eqref{energy(r)} does not depend on the inertial masses of the strings, but only on the string tension $1/\alpha'$.
This can be explained with a rather elementary observation.
Consider the behavior of the gravitational potential created by a generic extended source of size $l$. At distances much larger than $l$, the source is effectively point-like and the potential is proportional to the inertial mass.
But at distances much smaller than $l$, the source appears infinitely extended and the potential depends only on its mass density.
Furthermore, the decay of the potential is slower in this region.
In our case the D-brane backreaction stretches the open strings up to an infinite length.
Consequently the open strings interact like infinitely extended objects: the gravitational potential energy depends on their tension rather than their inertial masses, and the decay of the potential energy is slower than what is expected for point-like objects.

In the evaluation of the Nambu-Goto action, we observe that the leading contribution at large $r$ to the gravitational potential energy \eqref{energy(r)} comes from the part of the string that is closer to the horizon (region 1 in Figure \ref{figBackreact}).
However most of the inertial mass of the strings is located further away from the horizon, potentially even outside the throat created by the D-branes (region 2 in Figure \ref{figBackreact}).
This massive part of the string also contributes to the potential energy.
For instance if the asymptotic space is flat this contribution goes like $1/ r^{7}$, as for a point-particle in ten dimensions.
It is clearly subdominant at large $r$.

Because of the infinite redshift at the horizon, the geodesic distance between the string and the stack goes to zero near the D-branes.
 Thus the leading term in the potential energy  \eqref{energy(r)} comes from a small-distance effect in the bulk, even if the coordinate distance $r$ is very large.
In this  computation it is crucial that we consider extended strings that stretch up to the horizon.
The potential energy between point-like particles in the bulk would also be affected by the D-brane backreaction, but there would be no dramatic enhancement of the gravitational force as the one we observe here.  Thus the surprising behavior of the potential energy \eqref{energy(r)} is the result of a stringy effect.


\section{Compactification of the transverse space}

So far we considered extremal D3-branes in a non-compact ten-dimensional background.
In order to get a bit closer to phenomenology, we now consider extremal D3-branes in a four-dimensional compactification of string theory.
Notice that the D3-brane backreaction creates an infinite throat. Thus the transverse manifold decompactificaties, and the Kaluza-Klein spectrum of the graviton is actually continuous.
Nevertheless the graviton admits a normalizable zero-mode \cite{Randall:1999vf}\cite{Verlinde:1999fy}.
For our purposes the main modification with respect to the previous computation is that the zero-mode of the graviton produces a new contribution to the potential energy:
\be\label{energy(r)Compact}
 E(r) = -\frac{G_N^{(4)}}{r} \left(mM +  \frac{V^{(6)}}{L^6} \frac{N_{F1}\sqrt{g_s N}}{\alpha'}  \right) + {\cal O}\left(\frac{1}{r^2}\right) \ee
where we introduced the four-dimensional Newton's constant $G_N^{(4)} =  \alpha'^4 g_s^2/V^{(6)} $ with $V^{(6)}$ the volume of the transverse space and we have discarded numerical factors.
The first term in \eqref{energy(r)Compact} is the contribution of the graviton zero mode.
It is proportional to the inertial masses of the open string and of the open string stack (denoted respectively $m$ and $M$)\cite{Benichou:2010sj}: this is the usual four-dimensional Newtonian potential.
The second term in \eqref{energy(r)Compact} is the stringy contribution computed previously.

Let us determine which term dominates the potential energy \eqref{energy(r)Compact}. The ratio between the first term and the second term can be written as: $m(M/N_{F1})\alpha'\times L^6/V^{(6)} \times (1/\sqrt{g_s N})$.
The first factor $m(M/N_{F1})\alpha'$ involves the masses of the elementary particles in stringy units. In a phenomenological context this factor is much smaller than one.
Consider now the second factor $L^6/V^{(6)}$.
 A lower bound on the transverse volume $V^{(6)}$ is given by the sum of the volume of the throat $\sim L^6$ and the volume $V_{CY}$ of the compact transverse space without the D-branes: $V^{(6)} \gtrsim L^6 + V_{CY}$. Consequently the ratio $L^6/V^{(6)}$ is of order one or smaller. Finally for the field theory description of gravity to be reliable the third factor $1/\sqrt{g_s N}$ has to be smaller than one .
We conclude that the second term in \eqref{energy(r)Compact} dominates and the effective violation of the equivalence principle is large.

It should not come as a surprise that the Newtonian gravitational potential between fundamental strings receives corrections that are not proportional to inertial masses.
Indeed such corrections almost always appear when dealing with extended objects.
However they are expected to fade away when the distance between the objects is much larger than their typical size.
What is surprising in our case is that the stringy corrections to the Newtonian potential are visible at arbitrarily large distances.
The reason is that the backreaction of the extremal D3-branes stretches the open strings up to an infinite size.


\section{Turning on temperature}

We make one more step towards realistic string vacua by relaxing the condition that the D3-branes are extremal and allowing for a non-zero temperature.
In this case the throat created by the D3-branes has a finite length.
Consequently the open strings also have a finite proper length.
According to our previous  arguments, the stringy corrections to the Newtonian potential should fade away when the distance between the strings goes to infinity.

\begin{figure}[t]
\centering
\includegraphics[width=1\linewidth]{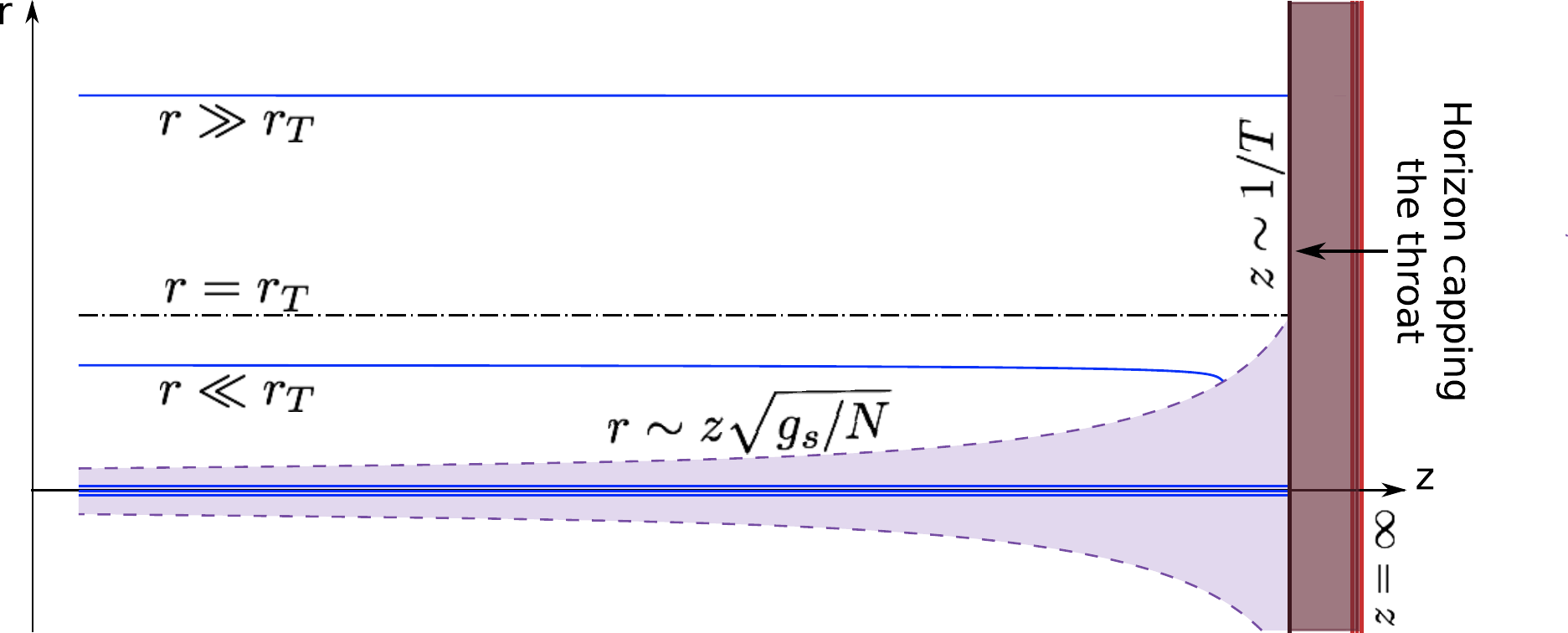}
\caption{When the temperature on the D3-branes is non-zero, the stringy corrections to the gravitational potential are visible for $r$ smaller than a critical value $r_T$.}
\label{figT}
\end{figure}

The computation of the gravitational potential between open strings in the full supergravity background for non-extremal D3 branes is beyond the scope of this work.
However using our previous results we can perform a qualitative analysis.
A finite temperature $T$ effectively caps off the AdS throat created by the D3-branes \eqref{metricAdS} at $z_T\sim 1/T$.
To take this effect into account we can simply put a cutoff on the geometry \eqref{metricWilsonLine} at $z=z_T$ (see Figure \ref{figT}).
The horizon created by the backreaction of the stack is located along the curve $r \sim z \sqrt{g_s/N}$ \cite{BE}.
We deduce that the behavior of the potential energy depends on whether $r$ is smaller or bigger than a critical length $r_T$ given by:
\be r_T = \frac{1}{T}\sqrt{\frac{g_s}{N}}\ee
For $r\ll r_T$, our previous analysis is essentially unchanged and the energy is given by (\ref{energy(r)Compact}).
For $r\gg r_T$  the stringy correction to the Newtonian potential fades: the open strings behave essentially as point-particles.

Let us  estimate the order of magnitude of the length $r_T$.
A temperature of one Kelvin corresponds to a distance of the order of a millimeter.
Consequently if $g_s$ is not too small, the distance $r_T$ up to which stringy effects are visible at very low temperature is much greater than the fundamental string scale and can even be macroscopic.
This suggests a new way to confront type II phenomenological models with experiment by testing the equivalence principle at very low temperature.


\section{Conclusions}

In general, D-brane backreaction causes a large increase in the open string proper length. Consequently stringy effects become visible at distances much greater than the elementary string scale.
As a concrete example we computed the gravitational potential  between open strings ending on D3-branes.
First we considered the case of extremal D3-branes in a non-compact spacetime.
We found that the D-brane backreaction greatly enhances the gravitational attraction: the potential goes like $1/r$, where $r$ is the distance measured by an observer living on the D-branes, even though gravity propagates in ten dimensions.
The dominant contribution to the potential comes from the section of the strings closest to the D-branes.
Then we discussed the case of a four-dimensional compactification of string theory, which is relevant for phenomenology.
We showed that the usual Newtonian potential receives a large stringy correction, which results in an apparent violation of the equivalence principle on the brane-world.
Finally we considered the effects of temperature which effectively caps the throat created by the D-branes.
We argued that the distance up to which the equivalence principle is strongly violated depends on the temperature.
It can be macroscopic when the temperature is low enough.
There is no direct tension with astrophysical and cosmological observations, as the CMB temperature of $3$K screens the correction down to scales of order a fraction of a millimeter.
However, one may hope to observe this effect in very low temperature tests of the equivalence principle.

\section*{Acknowledgments}
We would like to thank V. Hubeny and J. Troost for useful discussions.
We would also like to thank C. Bachas, C. Burgess, B. Craps and R. Russo for useful comments on a previous version of this letter.
This research is supported in part by the Belgian Federal Science Policy Office through the Interuniversity Attraction Pole Programme IAP VI/11-P.
R.B. is a Postdoctoral researcher of FWO-Vlaanderen, and is supported in part through project G011410N.
J.E. is supported by the FWO - Vlaanderen, Project No. G.0235.05.


\begin{thebibliography}{99}

\bibitem{BE}
  R.~Benichou and J.~Estes, ``Geometry of open strings ending on D3-branes,''
  to appear.


\bibitem{D'Hoker:2007fq}
  E.~D'Hoker, J.~Estes, M.~Gutperle,
  JHEP {\bf 0706 } (2007)  063.

\bibitem{Benichou:2010sj}
  R.~Benichou,
  JHEP {\bf 1010}, 001 (2010).

\bibitem{Randall:1999vf}
  L.~Randall, R.~Sundrum,
  Phys.\ Rev.\ Lett.\  {\bf 83 } (1999)  4690-4693.

\bibitem{Verlinde:1999fy}
  H.~L.~Verlinde,
  Nucl.\ Phys.\  {\bf B580 } (2000)  264-274.


\end{thebibliography}
\end{document}